\documentclass[%
 reprint,
superscriptaddress,
 amsmath,amssymb,
 aps,
]{revtex4-2}

\usepackage{graphicx}
\usepackage{dcolumn}
\usepackage{bm}
\usepackage{physics}
\usepackage{float}
\usepackage{multirow}
\usepackage{gensymb}
\usepackage{svg}
\usepackage{hyperref}
\usepackage[mathlines]{lineno}

\emergencystretch=\maxdimen
\begin{document}

\preprint{APS/123-QED}

\title{Experimental generation of polarization entanglement from spontaneous parametric down-conversion pumped by spatiotemporally highly incoherent light}

\author{Cheng Li}
\email{cli221@uottawa.ca}
\affiliation{Department of Physics, University of Ottawa, Ottawa, ON, Canada, K1N 6N5}

\author{Boris Braverman}
\affiliation{Department of Physics, University of Ottawa, Ottawa, ON, Canada, K1N 6N5}

\author{Girish Kulkarni}
\affiliation{Department of Physics, University of Ottawa, Ottawa, ON, Canada, K1N 6N5}

\author{Robert W. Boyd}
\email{rboyd@uottawa.ca}
\affiliation{Department of Physics, University of Ottawa, Ottawa, ON, Canada, K1N 6N5}
\affiliation{Institute of Optics, University of Rochester, Rochester, New York, USA, 14627}


\begin{abstract}
The influence of pump coherence on the entanglement produced in spontaneous parametric down-conversion (SPDC) is important to understand, both from a fundamental perspective, and from a practical standpoint for controlled generation of entangled states. In this context, it is known that in the absence of postselection, the pump coherence in a given degree of freedom (DOF) imposes an upper limit on the generated entanglement in the \textit{same} DOF. However, the cross-influence of the pump coherence on the generated entanglement in a \textit{different} DOF is not well-understood. Here, we experimentally investigate the effect of a spatiotemporally highly incoherent (STHI) light-emitting diode (LED) pump on the polarization entanglement generated in SPDC. Our quantum state tomography measurements using multimode collection fibers to reduce the influence of postselection yield a two-qubit state with a concurrence of \(0.531 \pm 0.006\) and a purity of \(0.647 \pm 0.005 \), in excellent agreement with our theoretically predicted concurrence of $0.552$ and purity of $0.652$. Thus, the use of an STHI pump leads to a reduction in the entanglement and purity of the output polarization two-qubit state. Nevertheless, the viability of SPDC with STHI pumps is important for two reasons: (i) STHI sources are ubiquitous and available at a wider range of wavelengths than lasers, and (ii) the generated STHI polarization-entangled two-photon states could potentially be useful in long-distance quantum communication schemes due to their robustness to scattering.
\end{abstract}

\maketitle



In the last few decades, entangled photon pairs produced from spontaneous parametric down-conversion (SPDC) \cite{Burnham1970PRL,Hong1985PRA,Boyd2020NLO} have become a ubiquitous resource for fundamental experiments in quantum optics \cite{Hong1987PRL,Ou1988PRL,Weihs1998PRL, Shih1988PRL,Kiess1993PRL, Wang1991PRA, Zou1991PRL} and practical realizations of quantum communication protocols \cite{Ursin2007NatPhysics,Yin2020Nature,Bouwmeester1997Nature}. In the context of SPDC, a number of studies have sought to understand the fundamental origin of the nonlocal correlations of the entangled photons, and how those correlations can be precisely tailored for various quantum applications \cite{Jha2010PRA,Giese2018PhysicaScripta,Monken1998PRA,Hugo2019PRA,Zhang2019OptExpress,Hutter2020PRL,Burlakov2001PRA,Jha2008PRA,Kulkarni2017JOSAB, Kulkarni2016PRA, Meher2020JOSAB}. In particular, the influence of the pump field's coherence properties on the generated two-photon entanglement has been investigated in the spatial \cite{Jha2010PRA,Giese2018PhysicaScripta,Monken1998PRA,Hugo2019PRA,Zhang2019OptExpress, Hutter2020PRL}, temporal \cite{Burlakov2001PRA,Jha2008PRA,Kulkarni2017JOSAB}, and polarization \cite{Kulkarni2016PRA, Meher2020JOSAB} degrees of freedom (DOFs). In each DOF, it was shown that, in the absence of postselection, the pump's coherence sets an upper bound on the generated entanglement in the same DOF. For instance, in the polarization DOF, if the setup is a closed system that does not involve postselection, the pump's polarization coherence determines the maximum achievable polarization entanglement of the generated two-qubit signal-idler state \cite{Kulkarni2016PRA}. However, such studies implicitly ignore or assume perfect pump coherence in every DOF other than the specific DOF under consideration. As a result, the cross-influence of the pump coherence in a given DOF on the entanglement generated in a different DOF is not well-understood. For instance, it is not well understood if a lack of spatial or temporal coherence in a perfectly-polarized pump field would prevent or somehow influence the two-qubit polarization entanglement between the signal and idler photons. In this context, a theoretical study predicts that polarization entanglement decreases with decreasing pump spatial coherence \cite{Sharma2021ResultsinPhysics}, whereas an experimental study concludes the exact opposite \cite{Ismail2017SciReports}. However, the conclusion of Ref.~\cite{Ismail2017SciReports} is dubious because their detection  employs single-mode fibers (SMFs) that effectively postselect for a spatially highly-coherent pump.

In this Letter, we experimentally investigate the polarization entanglement produced from SPDC pumped by a perfectly-polarized STHI beam from a light-emitting diode (LED). In contrast with Ref.~\cite{Ismail2017SciReports}, we collect the entangled photons using large-aperture multimode fibers (MMFs) to significantly reduce the influence of postselection. The output two-qubit state is measured to have a concurrence of \(0.531 \pm 0.006\) and a purity of \(0.647 \pm 0.005 \), in excellent agreement with our theoretically predicted concurrence of 0.552 and purity of 0.652. In essence, we find that the birefringence and dispersion of the crystal medium couple the spatiotemporal and polarization degrees of freedom in the experiment. Consequently, the lack of spatiotemporal coherence in the pump results in a degradation of the entanglement and purity of the output polarization two-qubit signal-idler state when measured in a non-postselective manner. In what follows, we describe the theory and then present the experimental results of our quantum state tomography and polarization correlation measurements.
\begin{figure}
\includegraphics[keepaspectratio]{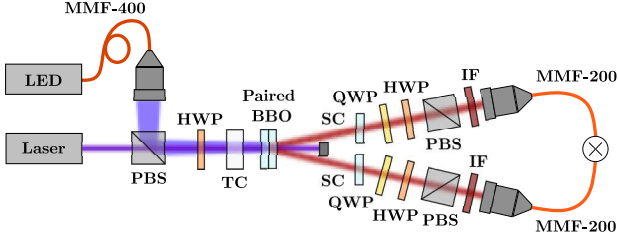}

\caption{\label{fig:1} Schematic diagram of the experimental setup. LED: light-emitting diode, PBS: polarizing beam splitter, HWP: half-wave plate, TC: temporal compensator, BBO: $\beta$-barium borate, SC: spatial compensator. QWP: quarter-wave plate. IF: interference filter. MMF-400 (200): multimode fiber with a core diameter of 400 (200) \(\mu\)m.}
\vspace{-1em}
\end{figure}

We consider SPDC pumped by spatiotemporally partially coherent light in the type-I double-crystal configuration outlined in Ref.~\cite{Kwiat1999PRA}. We define ${\bm q_{j}}$ and $\omega_{j}$ for $j=p,s,i$ as the transverse wave-vectors and frequencies corresponding to the pump, signal, and idler, respectively. From conservation of transverse momentum and energy, it follows that ${\bm q_{p}}={\bm q_{s}}+{\bm q_{i}}$ and $\omega_{p}=\omega_{s}+\omega_{i}$. The state vector describing the far-field spatiotemporal and polarization correlations of a given element in the generated two-photon ensemble can be written as (see Supplementary Material)
\begin{multline}\label{eqn1}
 |\psi\rangle=A\iiiint\mathrm{d}\bm q_{s}\mathrm{d}\bm q_{i}\mathrm{d}\omega_{s}\mathrm{d}\omega_{i}\,\Phi(\Delta k_{z}L)\\\times\Big[E_V({\bm q_{p}},\omega_p)|H,\bm q_{s},\omega_{s}\rangle|H,\bm q_{i},\omega_{i}\rangle\Big.\\\Big.+E_H({\bm q_{p}},\omega_p)e^{i\chi(\bm q_{s},\omega_{s},\bm q_{i},\omega_{i})}|V,\bm q_{s},\omega_{s}\rangle|V,\bm q_{i},\omega_{i}\rangle\Big],
\end{multline}
where $A$ is a normalization factor, $\Delta k_{z}$ is the longitudinal wavevector mismatch, $L$ is the length of each individual crystal, $\Phi(\Delta k_{z}L)=\mathrm{sinc}[\Delta k_{z}L/2]e^{i\Delta k_{z}L/2}$ is the phase-matching function, $E_{H(V)}({\bm q_{p}},\omega_p)$ denotes the horizontal (vertical) polarization component of the pump spectral amplitude inside the crystal, $|H(V),{\bm q_{j}},\omega_{j}\rangle$ for $j=s,i$ denotes the basis vector for the horizontal (vertical) polarization of the corresponding spatiotemporal mode of the signal and idler photon, respectively. The quantity $\chi(\bm q_{s},\omega_{s},\bm q_{i},\omega_{i})$ represents a relative phase acquired due to spatial and temporal walk-off between the two-photon state amplitudes generated in the two crystals. However, we assume that using temporal and spatial walk-off compensation elements in the setup, one can ensure $\chi(\bm q_{s},\omega_{s},\bm q_{i},\omega_{i})\approx 0$ (see Supplementary Material). For a pump field with spectral amplitude $E_{0}({\bm q_{p}},\omega_p)$ that is linearly polarized at $45^{\circ}$ or $-45^{\circ}$ before entering the crystal, we have $|E_{H}({\bm q_{p}},\omega_p)|=|E_{V}({\bm q_{p}},\omega_p)|=|E_{0}({\bm q_{p}},\omega_p)|/\sqrt{2}$ and the relative phase between $E_{H}({\bm q_{p}},\omega_p)$ and $E_{V}({\bm q_{p}},\omega_p)$ is $0$ or $\pi$, respectively. However, inside the crystal, the birefringence and dispersion properties of the medium induce a relative phase $\phi({\bm q_{p}},\omega_p)$ between $E_{H}({\bm q_{p}},\omega_p)$ and $E_{V}({\bm q_{p}},\omega_p)$. Using these relations, the final measured polarization two-qubit state $\rho$ can be written in the computational basis $\left\{|HH\rangle,|HV\rangle,|VH\rangle,|VV\rangle\right\}$ as
\begin{multline}\label{eqn2}
\rho=\mathrm{Tr}_{\mathrm{spat,temp}}(\langle|\psi\rangle\langle\psi|\rangle)=\begin{bmatrix}\,
   1/2 & 0 && 0 & \mu/2\\
   0 & 0 && 0 & 0\\
   0 & 0 && 0 & 0 \\
   \mu^*/2 & 0 && 0 & 1/2 
  \,\end{bmatrix},
\end{multline}
where $\mathrm{Tr}_{\mathrm{spat,temp}}$ denotes a partial trace over the spatial and temporal degrees of freedom, $\langle \cdots \rangle$ denotes an ensemble average over many realizations of the pump field, $|A|^2$ is chosen such that $\mathrm{Tr}(\rho)=1$. The quantity $\mu$ is then written as (see Supplementary Material)
\begin{equation}\label{eqn3}
\mu = |A|^2\int_{\Delta \bm q_p}\mathrm{d}{\bm q_{p}}\int_{\Delta\omega_p}\mathrm{d}\omega_{p}\ |E_{0}({\bm q_{p}},\omega_p)|^2\,e^{i\phi({\bm q_{p}},\omega_p)},
\end{equation}
where $\Delta \bm q_p = \Delta \bm q_{s}+\Delta \bm q_{i}$, and $\Delta \omega_p = \Delta\omega_{s}+\Delta\omega_{i}$; $\Delta \bm q_{s(i)}$ denotes the angular bandwidth corresponding to the detection aperture in the signal (idler) arm, $\Delta\omega_{s(i)}$ denotes the bandwidth of the spectral filter in the signal (idler) arm. It may be verified that $|\mu|$ satisfies $0\leq|\mu|\leq1$, with $|\mu|\to0$ when $\phi({\bm q_{p}},\omega_p)$ varies rapidly, and $|\mu|\to1$ when $\phi({\bm q_{p}},\omega_p)$ is  constant, in the integration region. Moreover, $|\mu|$ determines the purity $\mathrm{Tr}(\rho^2)=\{1+|\mu|^2\}/2$ and the concurrence $C(\rho)=|\mu|$ of the measured two-qubit state. In this work, we use concurrence to quantify the entanglement because in contrast with the Bell-CHSH parameter, which is only a sufficient but not necessary entanglement witness \cite{Werner1989PRA}, the concurrence can be used to quantify the degree of entanglement of an arbitrary two-qubit state \cite{Wootters1998PRL}.

\begin{figure*}[htp]
\includegraphics[width=0.98\textwidth,keepaspectratio]{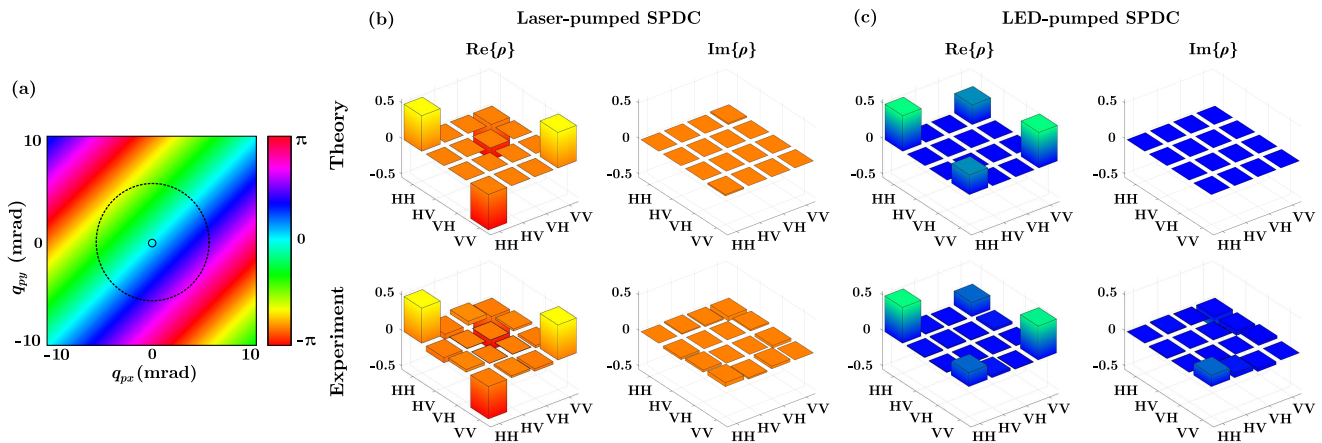}
\caption{\label{fig:2} (a) Theoretical 2D color plot of $\phi(\bm q_{p},\omega_{p0}=2\pi c/\lambda_{p0})$ for $\beta$-barium borate (BBO), where $\lambda_{p0}=405$ nm. The solid and dashed circles indicate the effective angular bandwidths of laser and LED pumps, respectively. (b) and (c) depict the theoretically predicted and experimentally measured density matrices for laser-pumped and LED pumped SPDC, respectively. Note that the laser pump and the LED pump at the PBS output are linearly polarized at $-45^\circ$ and $45^\circ$, respectively, because they are coupled into the crystal from different ports of the PBS as shown in Fig.~\ref{fig:1}.}
\vspace{-1.5em}
\end{figure*}

We now consider the effect of the pump's spatiotemporal coherence on the measured two-qubit state. For a spatiotemporally highly coherent pump such as a laser, the function $E_{0}({\bm q_{p}},\omega_p)$ is highly narrow. Consequently, the averaging of $e^{i\phi({\bm q_{p}},\omega_p)}$ in Eq.~(\ref{eqn3}) effectively occurs over a very small integration region, which leads to $|\mu|$ being close to unity, irrespective of $\Delta \bm q_{s(i)}$ and $\Delta \omega_{s(i)}$. This predicts high measured values of concurrence and purity of two-qubit states produced from a laser pump, regardless of the detection system. However, for an STHI pump such as an LED, the function $E_{0}({\bm q_{p}},\omega_p)$ has support over a broad range of $q_p$ and $\omega_p$, and consequently, the role of $\Delta \bm q_{s(i)}$ and $\Delta \omega_{s(i)}$ in the integrations in Eq.~(\ref{eqn3}) becomes significant. For a detection system that employs SMFs and narrow-band spectral filters, $\Delta \bm q_{s(i)}$ and $\Delta \omega_{s(i)}$ are small. As a result, the averaging of $e^{i\phi({\bm q_{p}},\omega_p)}$ in Eq.~(\ref{eqn3}) again effectively occurs over a small integration region constrained by the relations ${\bm q_{p}}={\bm q_{s}}+{\bm q_{i}}$ and $\omega_{p}=\omega_{s}+\omega_{i}$. This leads to a high value of $|\mu|$, implying high values of purity and concurrence. In other words, the detection postselection obscures the effect of the pump's lack of coherence.  However, for a detection system that employs MMFs and broad-band spectral filters, $\Delta \bm q_{s(i)}$ and $\Delta \omega_{s(i)}$ are large. Consequently, the averaging of $e^{i\phi({\bm q_{p}},\omega_p)}$ in Eq.~(\ref{eqn3}) occurs over a much larger range, and the effect of postselection is significantly reduced. In this case, Eq.~(\ref{eqn3}) would yield a reduced value of $|\mu|$, implying a degradation in the purity and concurrence of the measured state. Thus, our theory predicts that an STHI pump results in a degradation of the entanglement and purity of the output polarization two-qubit state, but to observe this degradation, it is necessary to use MMFs to collect the entangled photons.

In Fig.~\ref{fig:1}, we depict our experimental setup. The optical field from a Thorlabs M405L3 LED of center wavelength 405 nm, and full width at half maximum (FWHM) bandwidth of 20 nm, is butt-coupled into an MMF with core diameter 400 $\mu$m and numerical aperture 0.39, and coupled out into free space using a microscope objective. The LED light at the output of the fiber is measured to be 6 mW. The resulting collimated LED beam is then made linearly polarized at $45^{\circ}$ by passing it through a polarizing beam splitter (PBS) and a half-wave plate (HWP), and then made incident onto a paired $\beta$-barium borate (BBO) double-crystal cut for non-collinear emission with a half-opening angle of $3^{\circ}$ for perpendicular pump incidence. The double-crystal consists of two identically cut 0.5-mm-thick Type-I BBO crystals attached to each other with their optic axes oriented perpendicularly to one another. An ultraviolet continuous-wave laser with center wavelength 405 nm, bandwidth 2 nm, and power 20 mW is made linearly-polarized at $-45^{\circ}$ and aligned as pump for benchmarking purposes such that experiments performed with the LED pump can be compared with those performed with the laser pump for the same setup. In both cases, the pump photons are propagated through a 5-mm temporal compensation (TC) quartz crystal to pre-compensate for the temporal walk-off that the two orthogonal polarizations subsequently experience inside the double-crystal \cite{Nambu2002PRA,Rangarajan2009OptExpress}. The conjugate signal and idler photons from two diametrically opposite regions of the noncollinear emission ring are each sent through a 0.25-mm thick BBO spatial compensation (SC) crystal to compensate for spatial walk-off effects that the photons have experienced inside the double-crystal \cite{Altepeter2005OptExpress}. The photons are then passed through combinations of a quarter-wave plate (QWP), HWP, and PBS to measure their joint two-photon state. Note that we choose all the waveplates in the setup to be zero-order and the optics axes of all the waveplates are carefully aligned before taking measurements. This ensures that our results are not influenced by spectral dispersion or manufacturing imperfections of the waveplates. The diameters of waveplates are chosen to be much larger than the beam size to avoid postselection. The photons are sent through bandpass filters with center wavelength 810 nm and FWHM bandwidth of 10 nm. Subsequently, the photons are coupled into MMFs with core diameter 200 $\mu$m and numerical aperture 0.39 placed in the far-field. Finally, the photons from the MMFs are detected using PerkinElmer SPCM-QRH-14-FC avalanche photodiodes and their coincidence count rates are extracted using the Universal Quantum Devices (UQD) Logic-16 data-acquisition unit with a  coincidence time resolution window of one nanosecond.

We record coincidence rates for different rotation angles of the QWP and HWP in each arm, both for performing quantum state tomography (QST) of the two-photon polarization state, and measuring polarization correlation fringes in mutually unbiased bases. For the former purpose of QST, we record the coincidence rates with 16 different projective measurements, subtract the theoretical accidental count rates, and use the maximum likelihood state estimation (MLSE) algorithm outlined in Ref.~\cite{James2001PRA} to infer the two-qubit density matrix $\rho$. We then calculate the concurrence $C(\rho)$ \cite{Wootters1998PRL} and purity $\mathrm{Tr}[\rho^2]$. For the latter purpose of recording polarization correlation fringes, we fix the polarization in the signal arm and record the coincidence rate as a function of the linear polarization in the idler arm, which is defined by an angle \(\theta\) with respect to the H-polarization. For laser-pumped SPDC, we set an acquisition time of 10 s for each basis and the maximum coincidence rate is \(\sim 700 \ \text{s}^{-1}\). In contrast, for LED-pumped SPDC, we set an acquisition time of 60 min per basis, and the maximum coincidence rate was \(\sim 0.04\ \text{s}^{-1}\). Thus, the coincidence rate for LED-pumped SPDC is much smaller than that of laser-pumped SPDC, which could perhaps be explained by the following two reasons: (i) the LED has a lower power than the laser, and (ii) the down-conversion efficiency for the LED pump could be lower than that for the laser pump due to the lack of coherent phase-matching \cite{Zhao2020APLPhotonics}.

\begin{figure}
\includegraphics[width=0.48\textwidth,height=0.48\textheight,keepaspectratio]{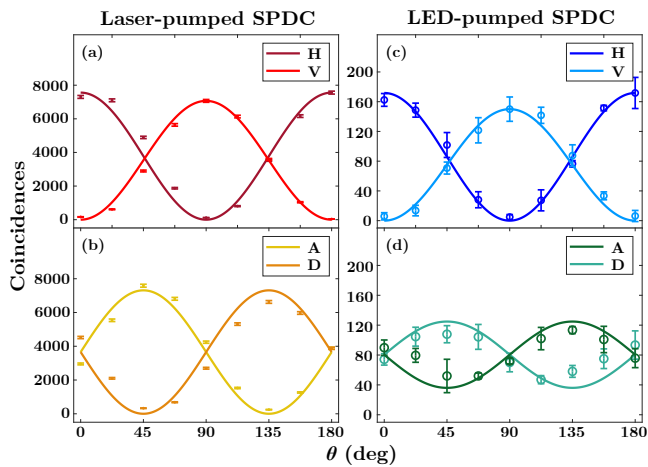}
\vspace{-1.5em}
\caption{\label{fig:3} Polarization correlation fringes in horizontal-vertical (H-V) basis and antidiagonal-diagonal (A-D) basis measured by fixing the signal photon's polarization and varying the idler photon's polarization represented by an angle $\theta$ with respect to the horizontal polarization. (a) and (b) correspond to laser-pumped SPDC, whereas (c) and (d) correspond to LED-pumped SPDC. The scatter plots represent the experimentally measured coincidence counts. The solid lines represent the theoretically predicted correlation fringes.}
\vspace{-1em}
\end{figure}

In Fig.~\ref{fig:2}(a), we depict the theoretically computed relative phase $\phi(\bm q_{p},\omega_{p})$. Since the relative phase has only negligible dependence on the wavelength of the pump (see Supplementary Material), we mainly focus on the dependence on the pump transverse wavevector and depict only the relative phase $\phi(\bm q_{p},\omega_{p0}=2\pi c/\lambda_{p0})$ for the central pump wavelength $\lambda_{p0}=405$ nm. Figs.~2(b) and 2(c) depict the theoretically predicted and experimentally measured two-photon density matrices for laser- and LED-pumped SPDC, respectively. For laser-pumped SPDC, the experimentally measured two-qubit state has a concurrence of \(0.955 \pm 0.003\) and purity of \(0.957 \pm 0.003\), in close agreement with the theoretically predicted concurrence $0.999$ and purity $0.999$. We verify that these results are highly stable over multiple measurements performed at different times. For LED-pumped SPDC, the experimentally measured two-qubit state has a concurrence of \(0.531 \pm 0.006\) and purity of \(0.647 \pm 0.005 \), in close agreement with the theoretically predicted concurrence of 0.552 and purity of 0.652. The fidelity of the experimentally measured density matrices to the theoretically predicted ones for laser-pumped and LED-pumped SPDC are $96.07$ \% and $94.92$ \%, respectively.

In Fig.~\ref{fig:3}, we depict the polarization correlation fringes in horizontal-vertical (H-V) and antidiagonal-diagonal (A-D) bases. Figs.~3(a) and 3(b) depict the case of laser-pumped SPDC, whereas 3(c) and 3(d) depict the case of LED-pumped SPDC. The scatter plots and solid lines represent the experimentally measured data points and the theoretically predicted correlation fringes, respectively. We notice that for laser-pumped SPDC, the fringes in  Figs.~3(a) and 3(b) exhibit high contrast in the H-V and A-D bases simultaneously. For quantitative characterization, we calculate the experimental fringe visibility values from sinusoidal fit curves of the measured counts. The fringe visibility in the H-V basis is \(98.8\pm1.2\%\), and the visibility in the A-D basis is \(93.7\pm1.8\%\). This presence of strong correlations in two mutually unbiased bases is the characteristic feature of entanglement. In contrast, for LED-pumped SPDC, the fringes in Fig.~3(c) corresponding to the H-V basis exhibit high contrast, but the fringes in Fig.~3(d) exhibit diminished contrast. In particular, the fringe visibility in the H-V basis is \(95.9\pm6.7\%\), but the fringe visibility in the A-D basis is \(39.4\pm3.9\%\). In other words, the polarization correlations in the H-V basis remain high, but those in the A-D basis are degraded, which is consistent with the entanglement reduction observed in our QST measurements.



In summary, we experimentally demonstrated the generation of polarization entanglement from SPDC pumped by an STHI LED source. We first presented a theoretical analysis that shows how the birefringence and dispersion of the crystal medium couple the spatiotemporal and polarization degrees of freedom in the experiment. This analysis predicts a degradation in the purity and entanglement of the two-qubit state produced from an STHI LED pump compared to that produced from a coherent laser pump. We then performed the experiment, both with an STHI LED pump and with a laser pump on the same setup for benchmarking purposes. In both cases, we showed that our experimental measurements of the two-qubit state using MMFs to reduce the influence of postselection are in excellent agreement with our theoretical predictions. For laser-pumped SPDC, the output two-qubit state has a concurrence of \(0.955 \pm 0.003\) and purity of \(0.957 \pm 0.003\), whereas for LED-pumped SPDC, the two-qubit state has a concurrence of \(0.531 \pm 0.006\) and a purity of \(0.647 \pm 0.005 \). Thus, the entanglement and purity of the polarization-entangled signal-idler state are lower for LED-pumped SPDC compared to laser-pumped SPDC. We also argue that setup instabilities cannot be the primary cause of this reduction because other studies employing highly-efficient periodically-poled crystals demonstrate that detection using SMFs yields high purity and entanglement despite the long acquisition times \cite{Zhang2022arXiv}. We were unable to perform such measurements using SMFs ourselves because of the poor efficiency of our bulk-crystal based SPDC setup.

In the future, our work might inspire further studies on SPDC with STHI sources. For instance, it may be possible to compensate for $\phi(\bm q_{p},\omega_{p})$ and enhance the polarization entanglement produced from a spatiotemporally incoherent pump. Moreover, the analysis in our study is restricted to type-I double-crystal SPDC, which employs critical phase-matching. However, there are also non-critical phase-matching methods that do not rely on birefringence, and therefore, do not couple the spatial and polarization degrees of freedom \cite{Boyd2020NLO,Wang2004PRA,Steinlechner2012OptExpress, Steinlechner2013OptExpress, Steinlechner2014JOSAB}. Using such methods, it may be possible to generate polarization two-qubit states with higher purity and entanglement with an STHI pump. Regardless, our study demonstrates the viability of using SPDC pumped by STHI sources for producing STHI polarization-entangled two-qubit states, which could have two important implications. Firstly, STHI sources such as LEDs and sunlight are ubiquitous and available at a wider range of wavelengths than their coherent counterparts such as lasers. Secondly, the STHI polarization-entangled two-qubit states produced from SPDC pumped by STHI sources might be well-suited for long-distance quantum communication schemes due to their robustness to scattering and turbulence \cite{Bhattacharjee2020OptLett,Qiu2012APB,Phehlukwayo2020PRA}.

During the course of this project, we became aware of a similar study being simultaneously carried
out by W. Zhang and coworkers \cite{Zhang2022arXiv}. We believe that our studies complement each other.

We acknowledge useful discussions with W. Zhang, E. Giese, J. Upham, J. Rioux and S. Lemieux. We also acknowledge funding from the Canada First Research Excellence Fund (Transformative Quantum Technologies). B.B. acknowledges support from the Banting Postdoctoral Fellowship. R.W.B. acknowledges support through the Natural Sciences and Engineering Research Council (NSERC) of Canada, the Canada Research Chairs program, by US DARPA award W911NF-18-1-0369, US ARO award W911NF-18-1-0337, US Office of Naval Research MURI award N00014-20-1-2558, US National Science Foundation Award 2138174, and a DOE award.

\nocite{*}

\bibliography{PRA_Pol}

\end{document}


\preprint{APS/123-QED}

\title{Supplementary material for: \\
Experimental generation of polarization entanglement from spontaneous parametric down-conversion pumped by spatiotemporally highly incoherent light}

\author{Cheng Li}
\email{cli221@uottawa.ca}
\affiliation{Department of Physics, University of Ottawa, Ottawa, ON, Canada, K1N 6N5}

\author{Boris Braverman}
\affiliation{Department of Physics, University of Ottawa, Ottawa, ON, Canada, K1N 6N5}

\author{Girish Kulkarni}
\affiliation{Department of Physics, University of Ottawa, Ottawa, ON, Canada, K1N 6N5}

\author{Robert W. Boyd}
\email{rboyd@uottawa.ca}
\affiliation{Department of Physics, University of Ottawa, Ottawa, ON, Canada, K1N 6N5}
\affiliation{Institute of Optics, University of Rochester, Rochester, New York, USA, 14627}



\maketitle

\section{Derivation of the two-photon density matrix in the polarization basis}

Consider a type-I double-crystal setup of the kind outlined in Ref.~\cite{Kwiat1999PRA}. We define ${\bm q_{j}}$ and $\omega_{j}$ for $j=p,s,i$ as the transverse wave-vectors and frequencies corresponding to the pump, signal, and idler, respectively. Due to energy and momentum conservation, we have $\bm{q}_p=\bm q_s+\bm q_i$ and $\omega_p=\omega_s+\omega_i$. In our analysis, the pump could be a general spatiotemporally partially coherent field, and consequently, the two-photon state generated from its SPDC will, in general, be a mixed state. Nevertheless, the final two-photon mixed state can be written as $\rho_\mathrm{spa,temp,pol}=\langle |\psi\rangle \langle\psi|\rangle$, where $\langle\cdots\rangle$ denotes an ensemble average over many realizations of the pump field, and each two-photon pure state element in the ensemble is written as
\begin{multline}\label{eqnS1.1}
 |\psi\rangle=A\iiiint\mathrm{d}\bm q_{s}\mathrm{d}\bm q_{i}\mathrm{d}\omega_{s}\mathrm{d}\omega_{i}\,\Phi(\Delta k_{z}L)\\\times\Big[E_V({\bm q_{p}},\omega_p)|H,\bm q_{s},\omega_{s}\rangle|H,\bm q_{i},\omega_{i}\rangle+E_H({\bm q_{p}},\omega_p)|V,\bm q_{s},\omega_{s}\rangle|V,\bm q_{i},\omega_{i}\rangle\Big],
\end{multline}
where $A$ is a normalization factor, $\Delta k_{z}$ is the longitudinal wavevector mismatch, $L$ is the length of each individual crystal, $\Phi(\Delta k_{z}L)=\mathrm{sinc}[\Delta k_{z}L/2]e^{i\Delta k_{z}L/4k_p}$ is the phase-matching function, $E_{H(V)}({\bm q_{p}},\omega_p)$ denotes the horizontal (vertical) polarization component of the pump spectral amplitude inside the crystal, $|H(V),{\bm q_{j}},\omega_{j}\rangle$ for $j=s,i$ denotes the basis vector for the horizontal (vertical) polarization of the corresponding spatiotemporal mode of the signal and idler photon, respectively. Since the pump polarization are always set at $45^{\circ}$ or $-45^{\circ}$, we can take $|E_{H}({\bm q_{p}},\omega_p)|=|E_{V}({\bm q_{p}},\omega_p)|$. Note that compared to Eq.~(1) in the main text, here we have dropped the term $e^{i\chi(\bm q_{s},\omega_{s},\bm q_{i},\omega_{i})}$, where the quantity $\chi(\bm q_{s},\omega_{s},\bm q_{i},\omega_{i})$ represents a relative phase acquired due to spatial and temporal walk-off between the two-photon state amplitudes generated in the two crystals. This approximation of $\chi(\bm q_{s},\omega_{s},\bm q_{i},\omega_{i})\approx 0$ is valid only if the spatial and temporal compensation schemes employed are adequate.

In order to verify the assumed adequateness of our compensation schemes, we note the following: In our experiment, the laser has a longitudinal coherence length ($\sim 150\ \mu \mathrm{m}$), which is shorter than the length of the crystal ($\sim 500\ \mu \mathrm{m}$ for each one of the double-crystal). Moreover, we employ multi-mode fibers (MMFs) instead of single-mode fibers (SMFs) to collect the down-converted photons in order to reduce spatial post-selection. Without adequate compensation, these experimental conditions cause temporal and spatial decoherence, which lead to degradation of entanglement and purity of the generated two-qubit state \cite{Altepeter2005OptExpress,Rangarajan2009OptExpress}. However, the fact that we measure two-qubit states with very high concurrence and purity for laser-pumped SPDC implies that our compensation schemes are adequate.

The two-photon density matrix in the joint basis of spatial, temporal and polarization degree of freedom is written as 
\begin{multline}\label{eqnS1.2}
 \rho_\mathrm{spa,temp,pol}=|A|^2\iiiint\mathrm{d}\bm q_{s}\mathrm{d}\bm q_{i}\mathrm{d}\bm q_{s}'\mathrm{d}\bm q_{i}'\iiiint\mathrm{d}\omega_{s}\mathrm{d}\omega_{i}\mathrm{d}\omega_{s}'\mathrm{d}\omega_{i}'\ |\Phi(\Delta k_{z}L)|^2\\
 \times\Big[\langle E_V({\bm q_{p}},\omega_p)E_V^*({\bm q_{p}'},\omega_p')\rangle|H,\bm q_{s},\omega_{s}\rangle|H,\bm q_{i},\omega_{i}\rangle \langle H,\bm q_{s}',\omega_{s}'|\langle H,\bm q_{i}',\omega_{i}'|\Big.\\
 \Big. +\langle E_V({\bm q_{p}},\omega_p)E_H^*({\bm q_{p}'},\omega_p')\rangle|H,\bm q_{s},\omega_{s}\rangle|H,\bm q_{i},\omega_{i}\rangle \langle V,\bm q_{s}',\omega_{s}'|\langle V,\bm q_{i}',\omega_{i}'|\Big.\\
 \Big. +\langle E_H({\bm q_{p}},\omega_p)E_V^*({\bm q_{p}'},\omega_p')\rangle|V,\bm q_{s},\omega_{s}\rangle|V,\bm q_{i},\omega_{i}\rangle \langle H,\bm q_{s}',\omega_{s}'|\langle H,\bm q_{i}',\omega_{i}'|\Big.\\
 \Big. +\langle E_H({\bm q_{p}},\omega_p)E_H^*({\bm q_{p}'},\omega_p')\rangle|V,\bm q_{s},\omega_{s}\rangle|V,\bm q_{i},\omega_{i}\rangle \langle V,\bm q_{s}',\omega_{s}'|\langle V,\bm q_{i}',\omega_{i}'|\Big],
\end{multline}
where $\langle \cdots \rangle$ denotes an ensemble average over many realizations of the pump field. The two-photon density matrix $\rho$ in the polarization degree of freedom can be obtained by taking the partial trace of $\rho_\mathrm{spa,temp,pol}$ over the spatial and temporal degrees of freedom, i.e,
\begin{multline}\label{eqnS1.3}
    \rho=\rho_\mathrm{pol}=\mathrm{Tr}_\mathrm{spa,temp}(\rho_\mathrm{spa,temp,pol})=|A|^2\iiiint\mathrm{d}\bm q_{s}\mathrm{d}\bm q_{i}\mathrm{d}\bm q_{s}'\mathrm{d}\bm q_{i}'\iiiint\mathrm{d}\omega_{s}\mathrm{d}\omega_{i}\mathrm{d}\omega_{s}'\mathrm{d}\omega_{i}'\\
 \times\delta(\bm q_{s}-\bm q_{s}')\,\delta(\bm q_{i}-\bm q_{i}')\,\delta(\omega_s-\omega_s')\,\delta(\omega_i-\omega_i')\ |\Phi(\Delta k_{z}L)|^2\\
 \times\Big[\langle E_V({\bm q_{p}},\omega_p)E_V^*({\bm q_{p}'},\omega_p')\rangle|HH\rangle \langle H H|+\langle E_V({\bm q_{p}},\omega_p)E_H^*({\bm q_{p}'},\omega_p')\rangle|HH\rangle \langle V V|\Big.\\
 \Big.+\langle E_H({\bm q_{p}},\omega_p)E_V^*({\bm q_{p}'},\omega_p')\rangle|VV\rangle \langle H H| +\langle E_H({\bm q_{p}},\omega_p)E_H^*({\bm q_{p}'},\omega_p')\rangle|VV\rangle \langle VV|\Big].
\end{multline}
Using the transverse momentum and energy conservation relations $\bm q_{p}^{(')}=\bm q_{s}^{(')}+\bm q_{i}^{(')}$ and $\omega_{p}^{(')}=\omega_{s}^{(')}+\omega_{i}^{(')}$, respectively, we obtain
\begin{multline}\label{eqnS1.4}
    \rho=|A|^2\iint\mathrm{d}\bm q_{p}\mathrm{d}\bm q_{-}\iint\mathrm{d}\omega_{p}\mathrm{d}\omega_{-}\ |\Phi(\Delta k_{z}L)|^2\\
 \times\Big[\langle E_V({\bm q_{p}},\omega_p)E_V^*({\bm q_{p}},\omega_p)\rangle|HH\rangle \langle H H|+\langle E_V({\bm q_{p}},\omega_p)E_H^*({\bm q_{p}},\omega_p)\rangle|HH\rangle \langle V V|\Big.\\
 \Big.+\langle E_H({\bm q_{p}},\omega_p)E_V^*({\bm q_{p}},\omega_p)\rangle|VV\rangle \langle H H| +\langle E_H({\bm q_{p}},\omega_p)E_H^*({\bm q_{p}},\omega_p)\rangle|VV\rangle \langle VV|\Big],
\end{multline}
where we have changed to the integration variables $\bm q_{p}=\bm q_{s}+\bm q_{i}$, $\bm q_{-}=\bm q_{s}-\bm q_{i}$, $\omega_{p}=\omega_{s}+\omega_{i}$ and $\omega_{-}=\omega_{s}-\omega_{i}$. We note that in the integrand, the phase-matching term depends only on $\bm q_-$ and $\omega_-$ \cite{Law2004PRL, Exter2006PRA, Kulkarni17JOSAB}, so that $\int\mathrm{d}\bm q_{-}\int\mathrm{d}\omega_{-} |\Phi(\Delta k_{z}L)|^2$ will just result in a scaling factor, which can be absorbed into $|A|^2$ together with other scaling factors resulting from the change of integration variables. Now Eqn.~\eqref{eqnS1.4} is reduced to
\begin{multline}\label{eqnS1.5}
    \rho=|A|^2\int_{\Delta\bm q_p}\mathrm{d}\bm q_{p}\int_{\Delta\omega_p}\mathrm{d}\omega_{p}\\
 \times\Big[\langle E_V({\bm q_{p}},\omega_p)E_V^*({\bm q_{p}},\omega_p)\rangle|HH\rangle \langle H H|+\langle E_V({\bm q_{p}},\omega_p)E_H^*({\bm q_{p}},\omega_p)\rangle|HH\rangle \langle V V|\Big.\\
 \Big.+\langle E_H({\bm q_{p}},\omega_p)E_V^*({\bm q_{p}},\omega_p)\rangle|VV\rangle \langle H H| +\langle E_H({\bm q_{p}},\omega_p)E_H^*({\bm q_{p}},\omega_p)\rangle|VV\rangle \langle VV|\Big],
\end{multline}
where the integration regions $\Delta \bm q_p$ and $\Delta \omega_p$ are the angular and spectral bandwidth of the components of the pump beam that participate in the SPDC process, respectively. These terms are denoted by $\Delta\bm q_p=\Delta\bm q_s+\Delta\bm q_i$ and $\Delta\omega_p=\Delta\omega_s+\Delta\omega_i$, where $\Delta\bm q_{s(i)}$ and $\Delta\omega_{s(i)}$ correspond to the size of the collection apertures in the momentum space and the bandwidth of the spectral filters inserted into the signal and idler arms, respectively.

The terms $\langle E_V({\bm q_{p}},\omega_p)E_V^*({\bm q_{p}},\omega_p)\rangle$ and $\langle E_H({\bm q_{p}},\omega_p)E_H^*({\bm q_{p}},\omega_p)\rangle$ are simply the intensity of the corresponding spatiotemporal mode of the pump $E_0({\bm q_{p}},\omega_p)$ and they are written as
\begin{equation}\label{eqnS1.6}
    \langle E_V({\bm q_{p}},\omega_p)E_V^*({\bm q_{p}},\omega_p)\rangle=\langle E_H({\bm q_{p}},\omega_p)E_H^*({\bm q_{p}},\omega_p)\rangle=|E_0({\bm q_{p}},\omega_p)|^2/2.
\end{equation}

The terms $\langle E_V({\bm q_{p}},\omega_p)E_H^*({\bm q_{p}},\omega_p)\rangle$ and $\langle E_H({\bm q_{p}},\omega_p)E_V^*({\bm q_{p}},\omega_p)\rangle$ characterize the relative phase $\phi$ between the V- and H-polarized components of the same spatiotemporal mode of the pump $E_0({\bm q_{p}},\omega_p)$ and they can be used to describe the polarization of that specific spatiotemporal mode. Although the pump polarization are set at $45^\circ$ or $-45^\circ$, the V- and H-polarized components can acquire different phases after entering the BBO crystals due to birefringence and dispersion. As a result, the relative phase become dependent on the spatiotemporal mode of the pump:
\begin{equation}\label{eqnS1.7}
    \langle E_V({\bm q_{p}},\omega_p)E_H^*({\bm q_{p}},\omega_p)\rangle=\langle E_H({\bm q_{p}},\omega_p)E_V^*({\bm q_{p}},\omega_p)\rangle^*=|E_0({\bm q_{p}},\omega_p)|^2e^{i\phi({\bm q_{p}},\omega_p)}/2.
\end{equation}

Now we normalize and rewrite Eqn.~\eqref{eqnS1.5} in the matrix form as
\begin{equation}\label{eqnS1.8}
    \rho=
    \begin{bmatrix}\,
    1/2 && 0 && 0 && \mu/2 \\
    0 && 0 && 0 && 0 \\
    0 && 0 && 0 && 0 \\
    \mu^*/2 && 0 && 0 &&  1/2
  \,\end{bmatrix},
\end{equation}
where we have taken $|A|^2\int_{\Delta\bm q_{p}}\mathrm{d}\bm q_{p}\int_{\Delta\omega_p}\mathrm{d}\omega_{p}|E_{0}({\bm q_{p}},\omega_p)|^2 = 1$. The quantity $\mu$ is then written as 
\begin{equation}\label{eqnS1.9}
    \mu = |A|^2\int_{\Delta\bm q_{p}}\mathrm{d}\bm q_{p}\int_{\Delta\omega_{p}}\mathrm{d}\omega_{p}|E_{0}({\bm q_{p}},\omega_p)|^2e^{i\phi({\bm q_{p}},\omega_p)}.
\end{equation}
\begin{equation}
    C = \bigg||A|^2\int_{\Delta\bm q_{p}}\mathrm{d}\bm q_{p}\int_{\Delta\omega_{p}}\mathrm{d}\omega_{p}|E_{0}({\bm q_{p}},\omega_p)|^2e^{i\phi({\bm q_{p}},\omega_p)}\bigg|,
\end{equation}
It can be shown that $\mu$ satisfies $0\leq|\mu|\leq1$, with $|\mu|\to1$ when $\phi(\bm q_p,\omega_p)$ only has negligible variations over the entire integration region, and $|\mu|\to0$ when $\phi(\bm q_p,\omega_p)$ varies rapidly within the integration region. Moreover, for a two-photon state in the form of \eqref{eqnS1.8}, $\mu$ determines its concurrence $C(\rho)=|\mu|$ and purity $\mathrm{Tr}(\rho^2)=\{1+|\mu|^2\}/2$.  

\begin{figure}
\includegraphics[width=0.68\textwidth,height=0.68\textheight,keepaspectratio]{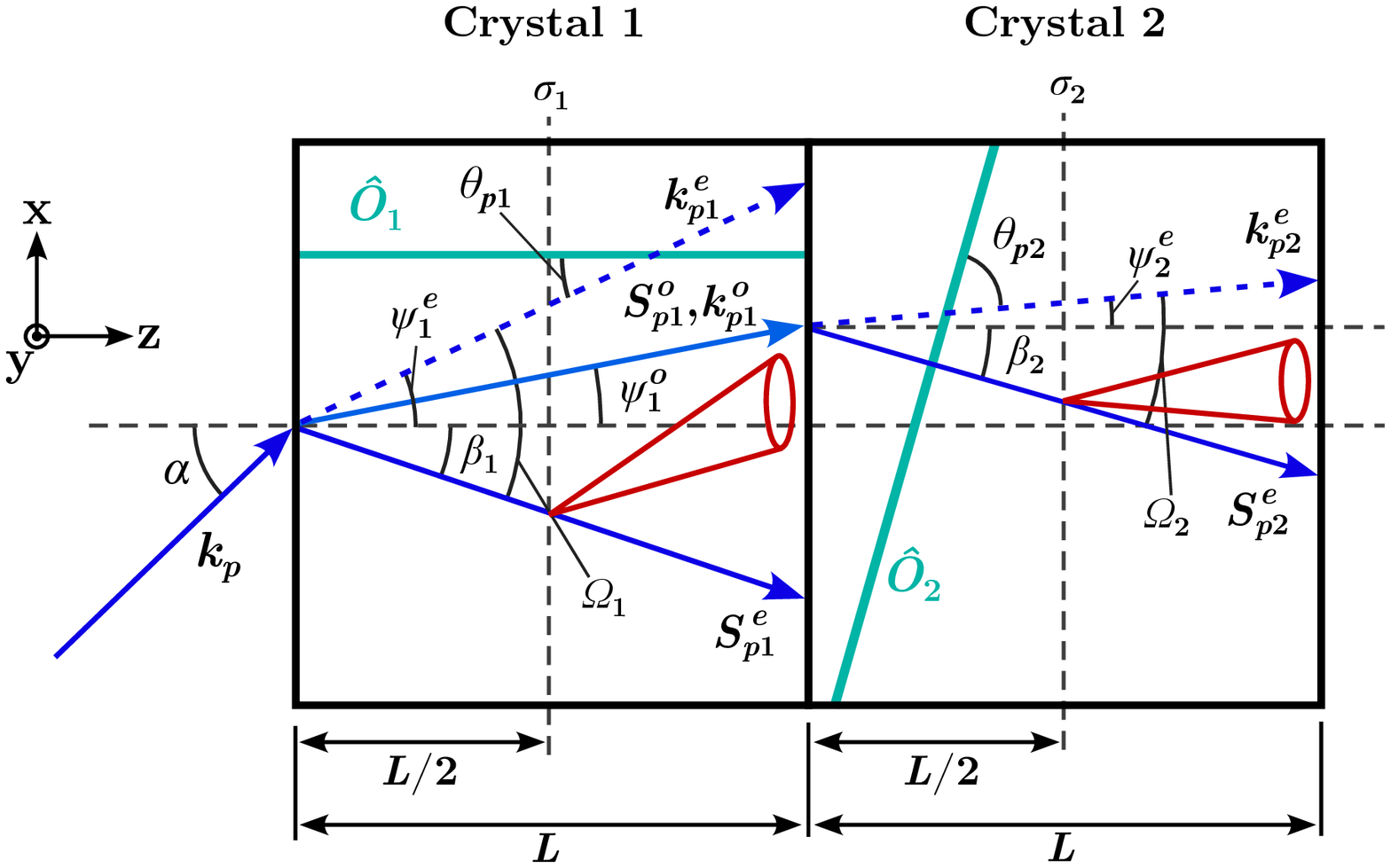}
\caption{\label{fig:S1} Schematic diagram illustrating the theoretical model used to calculate relative phase $\phi(\bm{q}_p,\omega_p)$. }
\end{figure}

\section{Derivation of relative phase}

In Fig.~\ref{fig:S1} we depict the theoretical model used to derive the expression for $\phi(\bm{q}_p,\omega_p)$. This model is adapted from the one proposed in Ref.~\cite{Altepeter2005OptExpress} and we expand it into the case of a pump beam with arbitrary incident angle. The paired-BBO crystal is placed in the x-y plane. The phase-matching condition denotes that in crystal 1, V-polarized component of the pump beam produces H-polarized PDC photon pairs; and in crystal 2, vice-versa. The relevant variables in this derivation are defined as follows: 

$\bm{k}_p$ -- Momentum vector of the pump beam outside of the crystal. 

$\bm{q}_p$ -- Transverse momentum vector of the pump beam outside of the crystal. Namely, $\bm{q}_p=k_{px}\hat{\bm{x}}+k_{py}\hat{\bm{y}}$.

$\hat{\bm{O}}_{1(2)}$ -- Unit vector for the optics axis of crystal 1 (crystal 2).

$\bm{k}_{p1(2)}^{e}$ -- Momentum vector of the \textbf{extraordinary} polarized pump beam inside crystal 1 (crystal 2).

$\bm{k}_{p1}^{o}$ -- Momentum vector of the \textbf{ordinary} polarized pump beam inside crystal 1.

$\bm{S}_{p1(2)}^{e}$ -- Poynting vector of the \textbf{extraordinary} polarized pump beam inside crystal 1 (crystal 2).

$\bm{S}_{p1}^{o}$ -- Poynting vector of the \textbf{ordinary} polarized pump beam inside crystal 1.

$\alpha$ -- Angle between $\bm{k}_p$ and $\hat{\bm{z}}$.

$\gamma$ -- Angle between $\bm{q}_p$ and $\hat{\bm{x}}$.

$\theta_{p1(2)}$ -- Angle between $\bm{k}_{p1(2)}^{e}$ and $\hat{\bm{O}}_{1(2)}$. 

$\psi_{p1(2)}^e$ -- Angle between $\bm{k}_{p1(2)}^{e}$ and $\hat{\bm{z}}$.

$\psi_{p1}^o$ -- Angle between $\bm{S}_{p1}^{o}$, $\bm{k}_{p1}^{o}$ and $\hat{\bm{z}}$.

$\beta_{1(2)}$ -- Angle between $\bm{S}_{p1(2)}^{e}$ and $\hat{\bm{z}}$.

$\Omega_{1(2)}$ -- Angle between $\bm{S}_{p1(2)}^{e}$ and $\bm{k}_{p1(2)}^{e}$.

$\Theta$ -- Angle between $\hat{\bm{O}}_{1(2)}$ and the x-y plane.
This quantity is the same for both crystal since they are cut the same way.

$L$ -- Length of crystal 1 and crystal 2.

$\sigma_{1(2)}$ -- The plane across the middle point of crystal 1 (crystal 2).

$\Phi_{1(2)}$ -- The phase accumulated by the V(H)-polarized pump beam in crystal 1(2) until reaching $\sigma_{1(2)}$.

$n_{e(o)}$ -- Refractive index of the extraordinary (ordinary) beam.

According to the laws of refraction and birefringence, some of the above variables will have the following relations \cite{Castelletto2004NJP}:
\begin{gather}
    \label{eqnS2.2}\psi_{p1}^o = \sin^{-1}\Big[\frac{\sin{\alpha}}{n_o(\omega_p)}\Big]\\
    \label{eqnS2.3}\psi_{p1}^e = \sin^{-1}\Big[\frac{\sin{\alpha}}{n_e(\theta_1,\omega_p)}\Big]\\
    \label{eqnS2.4}n_e(\theta_1,\omega_p) = n_o(\omega_p)\sqrt{\frac{1+\tan^2{\theta_1}}{1+[n_o(\omega_p)\tan{\theta_1}/\Bar{n_e}(\omega_p)]^2}}\\
    \label{eqnS2.5}\Omega_1 = \theta_1-\tan^{-1}\Big[\frac{n_o^2(\omega_p)}{\Bar{n_e}^2(\omega_p)}\tan(\theta_1)\Big]\\
    \label{eqnS2.6}\beta_1 = \Omega_1 - \psi_{p1}^e\\
    \label{eqnS2.7}\psi_{p2}^e = \sin^{-1}\Big[\frac{n_o(\omega_p)\sin{\psi_{p1}^o}}{n_e(\theta_2,\omega_p)}\Big]\\
    \label{eqnS2.8}n_e(\theta_2,\omega_p) = n_o(\omega_p)\sqrt{\frac{1+\tan^2{\theta_2}}{1+[n_o(\omega_p)\tan{\theta_2}/\Bar{n_e}(\omega_p)]^2}}\\
    \label{eqnS2.9}\Omega_2 = \theta_2-\tan^{-1}\Big[\frac{n_o^2(\omega_p)}{\Bar{n_e}^2(\omega_p)}\tan(\theta_2)\Big]\\
    \label{eqnS2.10}\beta_2 = \Omega_2 - \psi_{p2}^e
\end{gather}
In addition, the vectors for pump beam momentum and optic axes are written in the Cartesian coordinate representation as:
\begin{gather}
    \label{eqnS2.10}\hat{\bm{k}}_p = (\sin{\alpha}\cos{\gamma},\sin{\alpha}\sin{\gamma},\cos{\alpha})\\
    \label{eqnS2.11}\hat{\bm{k}}_{p1}^e = (\sin{\psi_{p1}^e}\cos{\gamma},\sin{\psi_{p1}^e}\sin{\gamma},\cos{\psi_{p1}^e})\\
    \label{eqnS2.12}\hat{\bm{k}}_{p2}^e = (\sin{\psi_{p2}^e}\cos{\gamma},\sin{\psi_{p2}^e}\sin{\gamma},\cos{\psi_{p2}^e})\\
    \label{eqnS2.13}\hat{\bm{O}}_1 = (0,\cos{\Theta},\sin{\Theta})\\
    \label{eqnS2.14}\hat{\bm{O}}_2 = (\cos{\Theta},0,\sin{\Theta})
\end{gather}
from the above expressions, we have
\begin{align}\nonumber
    \cos{\theta_1} &= \hat{\bm{k}}_{p1}^e\cdot\hat{\bm{O}}_1 \\\label{eqnS2.15}
    &= \sin{\psi_{p1}^e}\sin{\gamma}\cos{\Theta}+\cos{\psi_{p1}^e}\sin{\Theta}\\
\nonumber
    \cos{\theta_2} &= \hat{\bm{k}}_{p2}^e\cdot\hat{\bm{O}}_2 \\\label{eqnS2.16}
    &= \sin{\psi_{p2}^e}\sin{\gamma}\cos{\Theta}+\cos{\psi_{p2}^e}\sin{\Theta}
\end{align}

Since the down-converted photons are equally likely to be produced anywhere inside the region illuminated by the pump beam, in average, we can assume them to be produced in the middle plane of each crystal ($\sigma_{1}$ and $\sigma_{2}$). 

Now $\phi(\bm{q}_p,\omega_p)$ is equivalent to the difference between $\Phi_1$ and $\Phi_2$, which is written as
\begin{gather}\label{eqnS2.17}
\phi(\bm{q}_p,\omega_p) = \Phi_1 - \Phi_2\\
\label{eqnS2.18}\Phi_1 = \frac{2\pi n_e(\theta_1,\omega_p)}{\lambda_p}\frac{\cos{\Omega_1}}{\cos{\beta_1}}\frac{L}{2}\\
\label{eqnS2.19}\Phi_2 = \frac{2\pi n_e(\theta_2,\omega_p)}{\lambda_p}\frac{\cos{\Omega_2}}{\cos{\beta_2}}\frac{L}{2}
\end{gather}

\begin{figure*}
\includegraphics[width=\textwidth,height=\textheight,keepaspectratio]{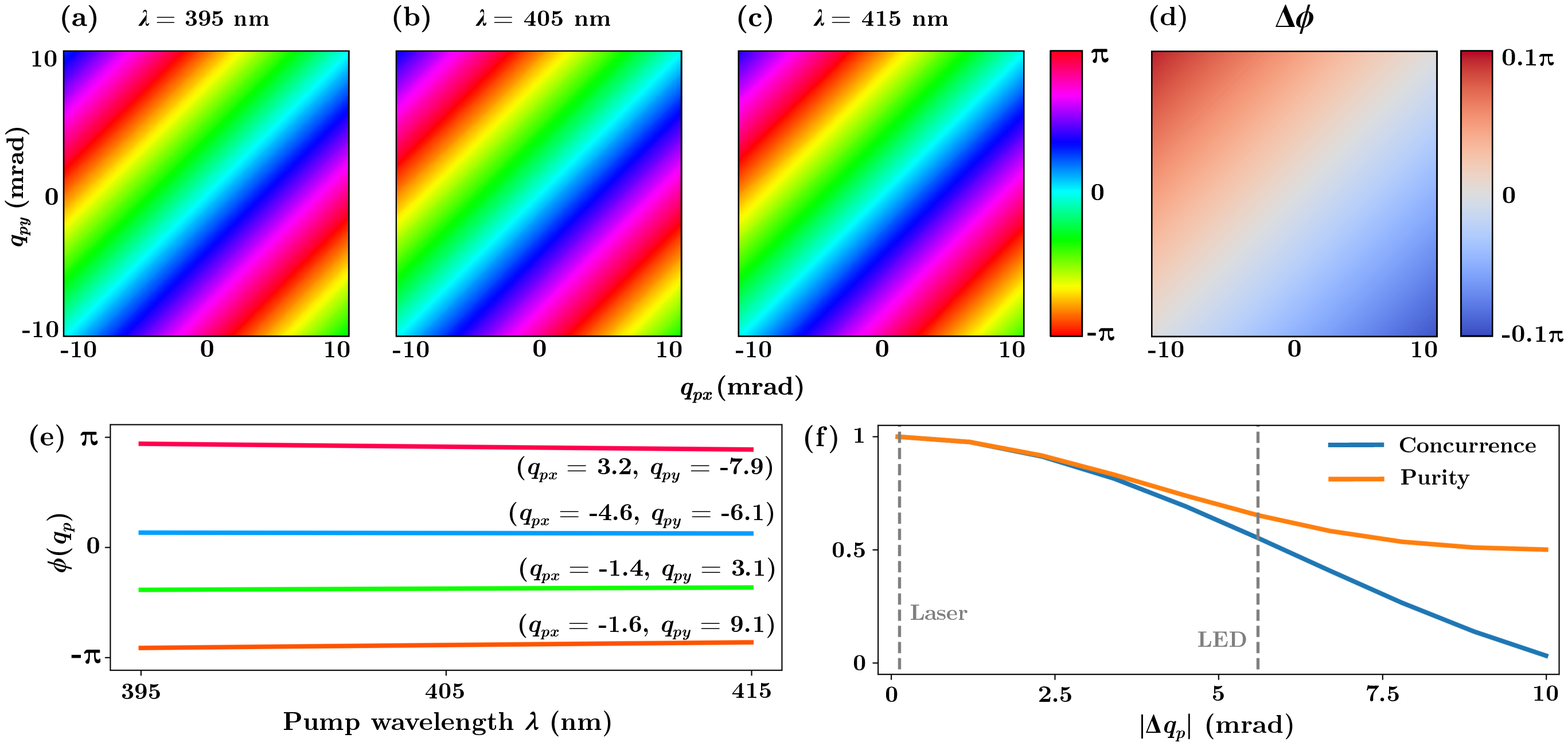}
\caption{\label{fig:S2} Theoretical 2D color plot of $\phi(\bm q_{p},\omega_{p}=2\pi c/\lambda_{p})$ for (a) $\lambda_{p}=395\ \mathrm{nm}$ (b) $\lambda_{p}=405\ \mathrm{nm}$ (c) $\lambda_{p}=415\ \mathrm{nm}$. (d) Theoretical 2D color plot of $\Delta \phi = \phi(\lambda_{p}=415\ \mathrm{nm}) - \phi(\lambda_{p}=395\ \mathrm{nm})$. (e) Theoretical plots of $\phi(\bm q_{p})$ at different pump wavelengths at four randomly chosen angular values. (f) Theoretical plots of concurrence and purity for different angular bandwidths of the pump. The grey dashed lines indicate the cases for laser and LED pump in our experiments.}
\end{figure*}

Now we can compute the dependence of $\phi(\bm{q}_p,\omega_p)$ on different spatiotemporal modes of the pump by varying the values of $\alpha$, $\gamma$, $\omega_p$. 

In Fig.~\ref{fig:S2}(a)-(c) we depict the relative phase $\phi(\bm{q}_p,\omega_p=2\pi c/\lambda_p)$ for different values of $\lambda_p$. In Fig.~\ref{fig:S2}(d) we depict the difference in $\phi$ between $\lambda_p = 395\ \mathrm{nm}$ and $\lambda_p = 415\ \mathrm{nm}$. In Fig.~\ref{fig:S2}(e) we depict the values of $\phi$ at different pump wavelengths for four randomly chosen angular values. It can be seen that the values of $\phi$ over the computed angular spectrum remain almost constant over the 20 nm wavelength bandwidth centered around 405 nm, with the maximum difference of $\phi$ being around $0.1\pi\ \mathrm{rad}$. This is because, for BBO crystals, the refractive index change introduced by birefringence is typically much larger than that introduced by dispersion near the 405 nm region. In Fig.~\ref{fig:S2}(f) we depict the values of concurrence and purity at different angular bandwidths of the pump. As shown by the plot, the concurrence and purity decrease as the angular bandwidth of the pump increases. In other words, averaging over different spatiotemporal modes with different phases reduces the polarization entanglement of the two-photon state.

Since it has been shown that $\phi$ is basically independent of $\omega_p$, we rewrite \eqref{eqnS1.9} as
\begin{equation}\label{eqnS2.20}
    \mu = |A|^2 \int_{\Delta\bm q_p}\mathrm{d}\bm q_{p}e^{i\phi({\bm q_{p}},\omega_{p0}=2\pi c/\lambda_{p0})},
\end{equation}
where $\lambda_{p0} = 405~\mathrm{nm}$ and by assuming that $E_{0}({\bm q_{p}},\omega_p)$ is slowly-varying over the integrated spatiotemporal bandwidth, $\int d\omega_p\ |E_{0}({\bm q_{p}},\omega_p)|^2$ results in a scaling factor that is absorbed into $|A|^2$. The integration region $\Delta\bm q_p$ is estimated according to the experimental conditions, as shown in the next section.

\section{Estimation of the effective angular bandwidth of the pump}

\begin{figure}
\includegraphics[width=0.58\textwidth,height=0.58\textheight,keepaspectratio]{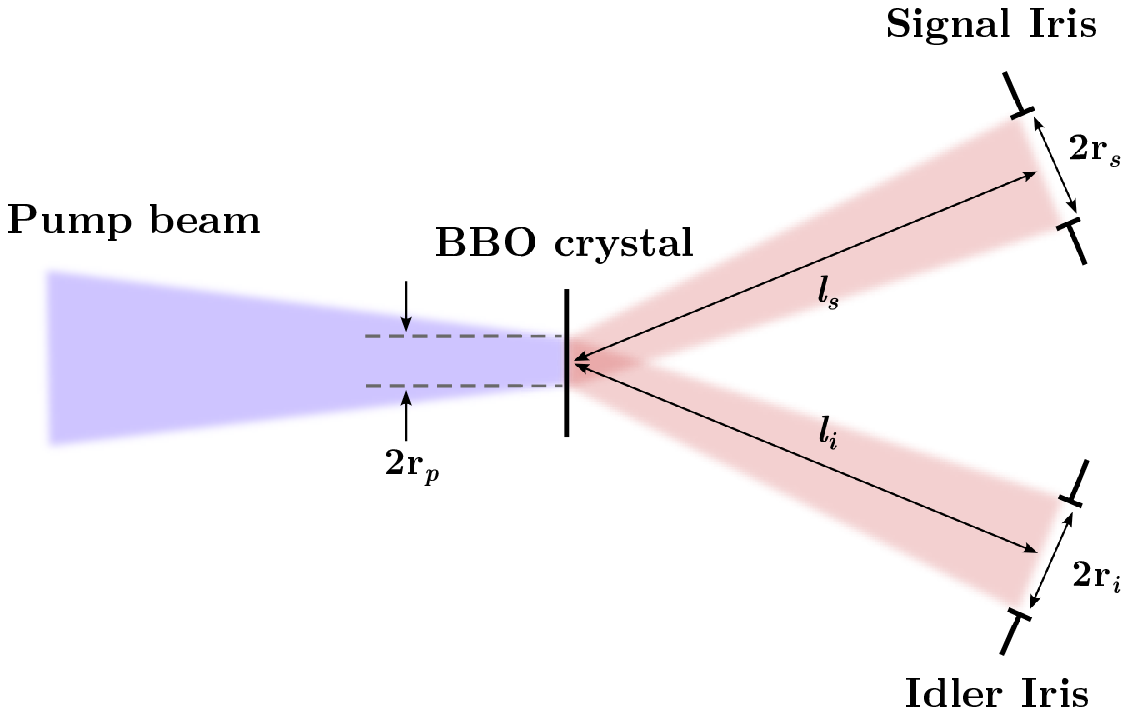}
\caption{\label{fig:S3} Schematic diagram illustrating the theoretical model used to estimate the effective angular bandwidth of the LED pump $\Delta\bm q_p$. }
\end{figure}

In Fig.~\ref{fig:S3} we depict an intuitive picture to estimate the effective angular bandwidth of the LED pump $\Delta\bm q_p$. We first place a variable iris in the signal arm and adjust the size until it barely blocks the down-conversion beam; we denote the radius of this iris as $\mathrm{r}_s$ and the distance between the crystal and the signal iris as $l_s$, respectively. Then the angular bandwidth of the signal beam can be estimated to be $|\Delta\bm q_s| = (\mathrm{r}_s-\mathrm{r}_p)/l_s$, where $\mathrm{r}_p$ is the diameter of the pump beam at the crystal plane. Similarly, to estimate the angular bandwidth of the idler beam, we have $|\Delta\bm q_s| = (\mathrm{r}_i-\mathrm{r}_p)/l_i$, where $\mathrm{r}_i$ and $l_i$ are the size of the iris on the idler arm and the distance between the crystal and the idler iris. The effective angular bandwidth of the pump can now be estimated as $|\Delta\bm q_p|\approx|\Delta\bm q_{s}|+|\Delta\bm q_{i}|$. 

In our setup, $|\Delta\bm q_p|$ is estimated to be 5.6 mrad. For a laser pump, one can calculate the integration upper limit by approximating it to the divergence half-angle of a Gaussian beam at the waist \cite{saleh2019fundamentals}, in our case, $|\Delta\bm q_p|_\mathrm{laser})$ is estimated to be 0.13 mrad.

\section{Differences between the theoretical and experimental results for LED-pumped SPDC}

\begin{figure}
\includegraphics[width=0.98\textwidth,height=0.98\textheight,keepaspectratio]{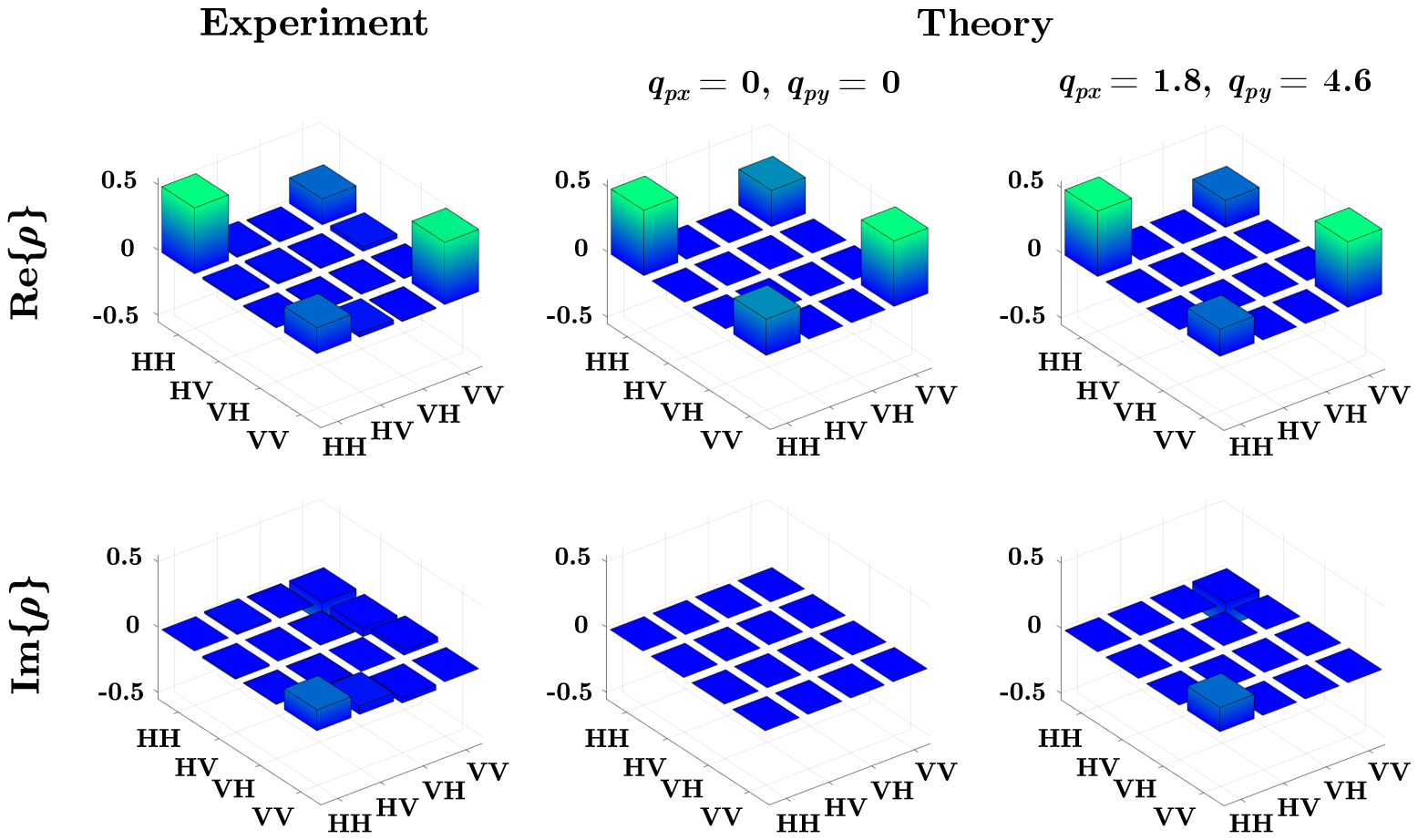}
\caption{\label{fig:S4} Experimentally measured and theoretically predicted density matrices for LED-pumped SPDC. The theoretically predicted ones are estimated with integration regions centered at different angular positions represented by $(\bm q_{px},\ \bm q_{py})$. The units for $\bm q_{px}$ and $\bm q_{py}$ are both mrad.}
\end{figure}

In the main text, we notice that the theoretically measured density matrix for LED-pumped SPDC contains non-zero off-diagonal elements in the imaginary part, while the theoretically predicted one does not. This can be explained by an alignment imperfection of the LED pump beam in the experimental setup. 

In producing the results shown in Fig.~2 of the main text, we estimated $\mu$ using
\begin{equation}\label{eqnS2.20}
    \mu = |A|^2 \int_{0<|\bm q_p|<|\Delta\bm q_p|}\mathrm{d}\bm q_{p}e^{i\phi({\bm q_{p}},\omega_{p0}=2\pi c/\lambda_{p0})},
\end{equation}

which assumes perfect alignment in the sense that the spatiotemporal modes of the pump beam that participate in the SPDC are centered around $(\bm q_{px} = 0, \bm q_{py} = 0)$. If the integration region is instead centered at $(\bm q_{px} = 1.8\ \mathrm{mrad}, \bm q_{py} = 4.6\ \mathrm{mrad})$, which corresponds to a 5 mrad angular deviation of the pump beam, we can obtain a density matrix that resembles the experimentally measured one. After accounting for the angular deviation, the theoretically predicted concurrence and purity are 0.553 and 0.653, respectively. Note that these results are close to those presented in the main text, which assumes perfect alignment and gives the theoretically predicted concurrence and purity as 0.552 and 0.652, respectively. The fidelity of the experimentally measured density matrix to the theoretically predicted one is now $97.92$ \% (see Fig.~\ref{fig:S4})

\bibliography{PRL_Pol_Supp}